\documentclass[10pt]{llncs}
\usepackage[latin1]{inputenc}
\usepackage[T1]{fontenc}
\usepackage{aeguill}
\usepackage{stmaryrd}
\usepackage{amssymb}
\usepackage{shortcuts}
\usepackage{xypic}
\usepackage{graphicx}
\usepackage{times}
\usepackage[all]{xy}

\renewcommand{\vxym}[1]{\vcenter{\xymatrix@C=2ex@R=2ex{#1}}}

\newcommand{\myparagraph}[1]{\smallskip\noindent\emph{\textbf{#1}}}

\newcommand{\G}{G}

\newcommand{\tile}[1]{\diamond_{#1}}
\newcommand{\initpos}[1]{{*}_{#1}}

\newcommand{\lbl}[2]{\ar@{}[#1]|-{#2}}
\newcommand{\hlbl}[1]{\lbl{#1}{\sim}}

\newcommand{\halting}[1]{\mathrm{halting}(#1)} 

\newcommand{\fixpoints}[1]{\mathrm{fix}(#1)} 
\newcommand{\domain}[1]{\mathrm{dom}(#1)} 

\newcommand{\cpos}[1]{{#1}^\circ} 

\newcommand{\booleangame}{\mathbb{B}} 
\newcommand{\questionmove}{\mathtt{q}} 
\newcommand{\answermove}{\mathtt{a}} 
\newcommand{\falsemove}{\mathtt{false}} 
\newcommand{\truemove}{\mathtt{true}} 
\newcommand{\questionposition}{q} 
\renewcommand{\root}{\ast} 
\newcommand{\tensorpair}[2]{#1\otimes #2} 
\newcommand{\opposant}[1]{#1} 
\newcommand{\joueur}[1]{#1} 

\newcommand{\before}{\varolessthan}
\newcommand{\after}{\varogreaterthan}

\title{Asynchronous games:\\innocence without alternation}

\author{Paul-Andr\'e Melli\`es \and Samuel Mimram\thanks{This work has been
    supported by the ANR Invariants alg\'ebriques des syst\`emes informatiques
    (INVAL). Physical address: Équipe PPS, CNRS and Universit\'e Paris~7, 2 place Jussieu,
    case 7017, 75251 Paris cedex 05, France. Email addresses:
    \hbox{\email{mellies@pps.jussieu.fr}} and
    \hbox{\email{smimram@pps.jussieu.fr}}.}}

\institute{}

\begin{document}
\maketitle

\pagestyle{plain}

\begin{abstract}
  The notion of innocent strategy was introduced by Hyland and Ong in order to
  capture the interactive behaviour of $\lambda$-terms and PCF programs. An
  innocent strategy is defined as an alternating strategy with partial memory,
  in which the strategy plays according to its view. Extending the definition to
  non-alternating strategies is problematic, because the traditional definition
  of views is based on the hypothesis that Opponent and Proponent alternate
  during the interaction.  Here, we take advantage of the diagrammatic
  reformulation of alternating innocence in asynchronous games, in order to
  provide a tentative definition of innocence in non-alternating games.  The
  task is interesting, and far from easy.  It requires the combination of true
  concurrency and game semantics in a clean and organic way, clarifying the
  relationship between asynchronous games and concurrent games in the sense of
  Abramsky and Melli\`es. It also requires an interactive reformulation of the
  usual acyclicity criterion of linear logic, as well as a directed variant, as
  a scheduling criterion.
\end{abstract}


\small
\section{Introduction}

\myparagraph{The alternating origins of game semantics.}
Game semantics was invented (or reinvented) at the beginning of the 1990s
in order to describe the dynamics of proofs and programs.
It proceeds according to the principles of trace semantics
in concurrency theory: every program and proof
is interpreted by the sequences of interactions, called \emph{plays},
that it can have with its environment.
The novelty of game semantics is that this set of plays defines a \emph{strategy}
which reflects the interactive behaviour of the program inside the \emph{game}
specified by the type of the program.
Game semantics was originally influenced by a pioneering work by Joyal~\cite{joyal:conway}
building a category of games (called Conway games) and alternating strategies.
In this setting, a game is defined as a decision tree (or more precisely, a dag) in which every edge,
called \emph{move}, has a polarity indicating whether it is played by the program,
called Proponent, or by the environment, called Opponent.
A play is alternating when Proponent and Opponent alternate strictly -- that is,
when neither of them plays two moves in a row.
A strategy is alternating when it contains only alternating plays.
The category of alternating strategies introduced by Joyal was later refined by
Abramsky and Jagadeesan~\cite{abramsky-jagadeesan:games-full-completeness} in
order to characterize the dynamic behaviour of proofs in (multiplicative) linear
logic.
The key idea is that the tensor product of linear logic, noted $\otimes$,
may be distinguished from its dual, noted $\llpar$,
by enforcing a \emph{switching policy} on plays -- ensuring
for instance that a strategy of~$A\otimes B$ reacts to an Opponent move
played in the subgame~$A$ by playing a Proponent move
in the same subgame~$A$.
The notion of \emph{pointer game} was then introduced by Hyland and Ong,
and independently by Nickau, in order to characterize
the dynamic behaviour of programs in the programming language PCF
-- a simply-typed $\lambda$-calculus extended with recursion,
conditional branching and arithmetical constants.
%
%
The programs of PCF are characterized dynamically as particular kinds
of strategy with partial memory -- called \emph{innocent}
because they react to Opponent moves according to
their own \emph{view} of the play.
This view is itself a play, extracted from the current play by removing all
its ``invisible'' or ``inessential'' moves.
This extraction is performed by induction on the length of the play,
using the pointer structure of the play, and the hypothesis
that Proponent and Opponent alternate strictly.

This seminal work on pointer games
led to the first generation of game semantics
for programming languages.
The research programme  -- mainly guided by Abramsky and his collaborators --
was extraordinarily successful: by relaxing in various ways the innocence constraint on strategies,
it suddenly became possible to characterize
the interactive behaviour of programs written in PCF (or in a call-by-value variant)
extended with imperative features like states, references, etc.
However, because Proponent and Opponent strictly alternate in the original definition
of pointer games, these game semantics focus on sequential languages
like Algol or ML, rather than on concurrent languages.
%


\myparagraph{Concurrent games.}
This convinced a little community of researchers to work
on the foundations of non-alternating games --
where Proponent and Opponent
are thus allowed to play several moves in a row at any point of the interaction.
Abramsky and
Melli\`es~\cite{abramsky-mellies:concurrent-games}
introduced a new kind of game semantics to that purpose, based on \emph{concurrent games}
-- see also \cite{abramsky:sequentiality-concurrency}.
In that setting, games are defined as partial orders (or more precisely, complete lattices)
of \emph{positions}, and strategies as \emph{closure operators} on these partial orders.
Recall that a closure operator~$\sigma$ on a partial order~$D$
is a function~$\sigma:D\longrightarrow D$ satisfying the following properties:
\begin{center}
\begin{tabular}{lclclcl}
(1) && $\sigma$ is increasing: &\hspace{1em} & $\forall x\in D,$ & \hspace{2em} & $x\leq \sigma(x)$,\\
(2) && $\sigma$ is idempotent: && $\forall x\in D,$ && $\sigma(x)=\sigma(\sigma(x))$,\\
(3) && $\sigma$ is monotone: &&  $\forall x,y\in D,$ && $x\leq y \Rightarrow \sigma(x)\leq \sigma(y)$.
\end{tabular}
\end{center}
The order on positions $x\leq y$ reflects the intuition that the position~$y$
contains more information than the position~$x$.
Typically, one should think of a position~$x$ as a set of moves in a game,
and $x\leq y$ as set inclusion~$x\subseteq y$.
Now, Property~(1) expresses that a strategy~$\sigma$ which transports
the position~$x$ to the position~$\sigma(x)$ increases the amount of
information.
Property~(2) reflects the intuition that the strategy~$\sigma$ delivers all its
information when it transports the position~$x$ to the position~$\sigma(x)$,
and thus transports the position~$\sigma(x)$ to itself.
Property~(3) is both fundamental and intuitively right, but also more subtle to justify.
Note that the interaction induced by such a strategy~$\sigma$ is possibly non-alternating,
since the strategy transports the position~$x$ to the position~$\sigma(x)$
by ``playing in one go'' all the moves appearing in~$\sigma(x)$ but not in~$x$.

\myparagraph{Asynchronous transition systems.}
Every closure operator~$\sigma$ is characterized by the set~$\fixpoints{\sigma}$
of its fixpoints, that is, the positions~$x$ satisfying~$x=\sigma(x)$.
%
%
So, a strategy is expressed alternatively 
as a set of positions (the set of fixpoints of the closure operator) in concurrent games,
and as a set of alternating plays in pointer games.
In order to understand how the two formulations of strategies
are related, one should start from an obvious analogy with concurrency theory:
pointer games define an \emph{interleaving semantics}
(based on sequences of transitions) whereas concurrent games define
a \emph{truly concurrent semantics} (based on sets of positions, or states)
of proofs and programs.
Now, Mazurkiewicz taught us this important lesson: a truly concurrent semantics
may be regarded as an interleaving semantics (typically a transition system)
equipped with \emph{asynchronous tiles} -- represented diagrammatically
as 2-dimensional tiles
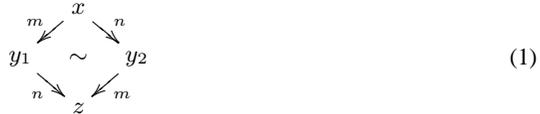
\begin{equation}
\label{equation/tuile}
\vxym{
  &\ar[dl]_{m}x\ar[dr]^{n}&\\
  y_1\ar[dr]_{n}&\sim&\ar[dl]^{m}y_2\\
  &z&\\
}
\end{equation}
expressing that the two transitions~$m$ and~$n$ from the state~$x$ are \emph{independent},
and consequently, that their scheduling does not matter
from a truly concurrent point of view.
%
%
This additional structure induces an equivalence relation
on transition paths, called \emph{homotopy},
defined as the smallest congruence relation~$\sim$
identifying the two schedulings $m\cdot n$ and $n\cdot m$ for every tile
of the form~(\ref{equation/tuile}).
The word \emph{homotopy} should be understood mathematically
as (directed) homotopy in the topological presentation of asynchronous
transition systems as \emph{$n$-cubical sets}~\cite{goubault:geometry-concurrency}.
This 2-dimensional refinement of usual 1-dimensional transition systems
enables to express simultaneously the interleaving semantics
of a program as the set of transition paths it generates,
and its truly concurrent semantics, as the homotopy classes
of these transition paths.
When the underlying 2-dimensional transition system is contractible
in a suitable sense, explained later, these homotopy classes coincide in fact
with the positions of the transition system.

\myparagraph{Asynchronous games.}
Guided by these intuitions, Melli\`es introduced the notion of \emph{asynchronous game},
which unifies in a surprisingly conceptual way the two heterogeneous notions
of pointer game and concurrent game.
Asynchronous games are played on asynchronous (2-dimensional)
transition systems, where every transition (or move) is equipped with
a \emph{polarity}, expressing whether it is played by Proponent or by Opponent.
A \emph{play} is defined as a path starting from the root (noted~$\ast$) of the game,
%
%
%
and a \emph{strategy} is defined as a well-behaved set of alternating plays,
in accordance with the familiar principles of pointer games.
Now, the difficulty is to understand how (and when) a strategy defined
as a set of plays may be reformulated as a set of positions, in the spirit
of concurrent games.

The first step in the inquiry is to observe that the asynchronous tiles~(\ref{equation/tuile})
offer an alternative way to describe \emph{justification pointers} between moves.
For illustration, consider the boolean game~$\booleangame$, where Opponent
starts by asking a question~$\questionmove$, and Proponent answers
by playing either~$\truemove$ or~$\falsemove$.
The game is represented by the decision tree
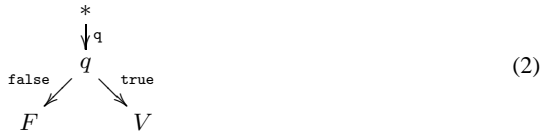
\begin{equation}\label{equation/booleangame}
\vcenter{\xymatrix @-1.2pc {
& \ast\ar[d]^-{\questionmove}\\
&
q
\ar[dr]^-{\truemove}
\ar[dl]_-{\falsemove}
\\
F && V
\\
}}
\end{equation}
where~$\ast$ is the root of the game, and the three remaining positions
are called~$q,F$ and $V$ ($V$ for ``Vrai'' in French).
At this point, since there is no concurrency involved, the game may be seen
either as an asynchronous game, or as a pointer game.
Now, the game~$\booleangame\otimes\booleangame$
is constructed by taking two boolean games ``in parallel.''
It simulates a very simple computation device,
containing two boolean memory cells.
In a typical interaction, Opponent starts by asking with~$\questionmove_L$
the value of the left memory cell, and Proponent answers~$\truemove_L$;
then, Opponent asks with~$\questionmove_R$ the value of the right memory cell,
and Proponent answers~$\falsemove_R$.
%
%
The play is represented as follows in pointer games:
\bigskip
$$
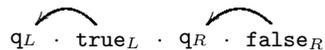
\xymatrix @-2pc {
\questionmove_L & \cdot & \truemove_L
\ar@/_1pc/[ll]
& \cdot &
\questionmove_R & \cdot & \falsemove_R
\ar@/_1pc/[ll]}
$$
The play contains two justification pointers, each one represented
by an arrow starting from a move and leading to a previous move.
Typically, the justification pointer from the move~$\truemove_L$
to the move~$\questionmove_L$ indicates that the answer~$\truemove_L$
is necessarily played after the question~$\questionmove_L$.
The same situation is described using 2-dimensional tiles
in the asynchronous game~$\booleangame\otimes\booleangame$
below:
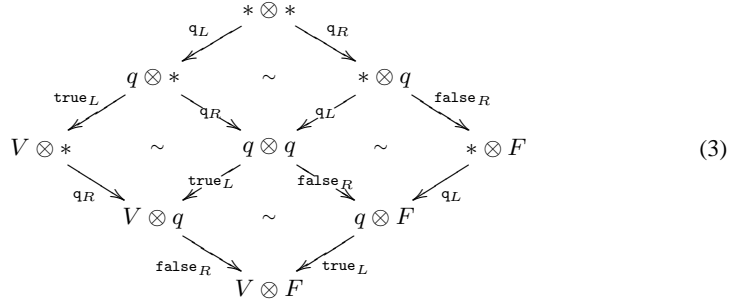
\begin{equation}
\label{equation/true-tensor-false}
\vcenter{\xymatrix @-1.7pc{
&&&&
\tensorpair{\root}{\root}
\ar[lldd]_{\opposant{\questionmove_L}}\ar[rrdd]^{\opposant{\questionmove_R}}\\
\\
&&
\tensorpair{\questionposition}{\root}
\ar[lldd]_{\joueur{\truemove_L}}\ar[rrdd]|-{\opposant{\questionmove_R}}
\ar@{}[rrrr]|-{\sim}
&& &&
\tensorpair{\root}{\questionposition}
\ar[lldd]|-{\opposant{\questionmove_L}} \ar[rrdd]^{\joueur{\falsemove_R}}
\\
\\
\tensorpair{V}{\root}\ar[rrdd]_{\opposant{\questionmove_R}}
\ar@{}[rrrr]|-{\sim}
&& && \tensorpair{\questionposition}{\questionposition}\ar[lldd]|-{\joueur{\truemove_L}}
\ar[rrdd]|-{\joueur{\falsemove_R}}
\ar@{}[rrrr]|-{\sim}
&& && \tensorpair{\root}{F}
\ar[lldd]^{\opposant{\questionmove_L}}\\
\\
&& \tensorpair{V}{\questionposition}\ar[rrdd]_{\joueur{\falsemove_R\ }}
\ar@{}[rrrr]|-{\sim}
&& &&
\tensorpair{\questionposition}{F}\ar[lldd]^{\joueur{\truemove_L}}\\
\\
&&&& \tensorpair{V}{F}
\\
}}
\end{equation}
%
%
%
The justification pointer between the answer~$\truemove_L$ and its question~$\questionmove_L$
is replaced here by a \emph{dependency} relation between the two moves, ensuring
that the move~$\truemove_L$ cannot be permuted before the move~$\questionmove_L$.
The dependency itself is expressed by a ``topological'' obstruction:
the lack of a 2-dimensional tile permuting the transition~$\truemove_L$
before the transition~$\questionmove_L$
in the asynchronous game~$\booleangame\lltens\booleangame$.

This basic correspondence between justification pointers and asynchronous tiles
allows a reformulation of the original definition of \emph{innocent strategy}
in pointer games (based on views) in the language of asynchronous games.
Surprisingly, the reformulation leads to a purely local
and diagrammatic definition of innocence in asynchronous games,
which does not mention the notion of view any more.
This diagrammatic reformulation leads then to the important discovery
that innocent strategies are \emph{positional} in the following sense.
Suppose that two alternating plays~$s,t:\ast\longrightarrow x$ 
with the same target position~$x$ are elements of an innocent strategy~$\sigma$,
and that~$m$ is an Opponent move from position~$x$.
Suppose moreover that the two plays~$s$ and~$t$ are equivalent modulo homotopy.
Then, the innocent strategy~$\sigma$ extends the play~$s\cdot m$
with a Proponent move~$n$ if and only if it extends the play~$t\cdot m$
with the same Proponent move~$n$.
Formally:
\begin{equation}\label{equation/positionality-alternating}
s\cdot m\cdot n \in \sigma
\hspace{1em}
\mbox{and}
\hspace{1em}
s\sim t
\hspace{1em}
\mbox{and}
\hspace{1em}
t\in\sigma
\hspace{1em}
\text{implies}
\hspace{1em}
t\cdot m\cdot n \in \sigma.
\end{equation}
From this follows that every innocent strategy~$\sigma$ is characterized
by the set of \emph{positions} (understood here as homotopy classes of plays)
reached in the asynchronous game.
This set of positions 
defines a closure operator, and thus
a strategy in the sense of concurrent games.
Asynchronous games offer in this way an elegant
and unifying point of view on pointer games and concurrent games.
\myparagraph{Concurrency in game semantics.}
There is little doubt that a new generation of game semantics
is currently emerging along this foundational work on concurrent games.
We see at least three converging lines of research.
First, authors trained in game semantics -- Ghica,
Laird and Murawski -- were able to characterize the interactive behaviour
of various concurrent programming languages like Parallel Algol~\cite{ghica-murawski:angelic-semantics}
or an asynchronous variant of the $\pi$-calculus~\cite{laird:pi-calc}
using directly (and possibly too directly) the language of pointer games.
Then, authors trained in proof theory and game semantics -- Curien and Faggian --
relaxed the sequentiality constraints required by Girard on \emph{designs} 
in ludics, leading to the notion of~$L$-net~\cite{curien-faggian:L-nets}
which lives at the junction
of syntax (expressed as proof nets) and game semantics
(played on event structures).
Finally, and more recently, authors trained in process calculi,
true concurrency and game semantics -- Varacca and Yoshida --
were able to extend Winskel's truly concurrent semantics of~CCS,
based on event structures, to a significant fragment of the $\pi$-calculus,
uncovering along the way a series of nice conceptual properties of \emph{confusion-free}
event structures~\cite{varacca-yoshida:typed-event-structures}.
%
%
%
%
%
%
%
%
%

%
So, a new generation of game semantics for concurrent programming languages
is currently emerging... but their various computational models are still poorly connected.
We would like a regulating theory here, playing the role of Hyland and Ong
pointer games in traditional (that is, alternating) game semantics.
Asynchronous games are certainly a good candidate, because they combine
interleaving semantics and causal semantics in a harmonious way.
Unfortunately, they were limited until now to alternating strategies~\cite{mellies:ag2}.
The key contribution of this paper is thus to extend
the asynchronous framework to non-alternating strategies in a smooth way.

\myparagraph{Asynchronous games without alternation.}
One particularly simple recipe to construct an asynchronous game
is to start from a \emph{partial order} of events where, in addition, every event has a polarity, indicating whether it is played by Proponent or Opponent.
This partial order~$(M,\preceq)$ is then equipped with a \emph{compatibility relation}
satisfying a series of suitable properties -- defining what Winskel
calls an \emph{event structure}.
A \emph{position}~$x$ of the asynchronous game is defined as a set of \emph{compatible}
events (or moves) closed under the ``causality'' order:
$$
\forall m,n\in M,  \hspace{2em} m\preceq n \hspace{1em} \mbox{and} \hspace{1em} n\in x
\hspace{1em}
\text{implies} \hspace{1em} m\in x.$$
Typically, the boolean game~$\booleangame$ described in (\ref{equation/booleangame})
is generated by the event structure
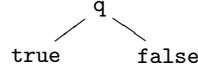
\begin{small}
$$
\xymatrix @-1.5pc {
&\questionmove
\ar@{-}[dl]\ar@{-}[dr]
\\
\truemove && \falsemove
}
$$
\end{small}

\noindent
where~$\questionmove$ is an Opponent move, and $\falsemove$ and $\truemove$
are two \emph{incompatible} Proponent moves, with the positions~$q,V,F$
defined as~$q=\{\questionmove\}$, $V=\{\questionmove,\truemove\}$
and $F=\{\questionmove,\falsemove\}$.
The tensor product~$\booleangame\otimes\booleangame$ of two boolean games
is then generated by putting side by side the two event structures, in the expected way.
The resulting asynchronous game looks like a flower with four petals,
one of them described in~(\ref{equation/true-tensor-false}).
More generally, every formula of linear logic defines an event structure --
which generates in turn the asynchronous game associated to the formula.
For instance, the event structure induced by the formula
\begin{equation}
\label{equation/b-tensor-b-pop-b}
(\booleangame\otimes\booleangame)\multimap\booleangame
\end{equation}
contains the following partial order of compatible events:
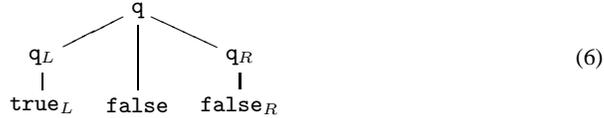
\begin{small}
\begin{equation}
\label{equation/slice-of-the-game}
\vcenter{\xymatrix @-1.5pc{
&& \questionmove\ar@{-}[dl]\ar@{-}[dr]\ar@{-}[dd]&&
\\
& \questionmove_L \ar@{-}[d] && \questionmove_R \ar@{-}[d]
\\
& \truemove_L & \falsemove & \falsemove_R
}}
\end{equation}
\end{small}

\noindent
which may be seen alternatively as a (maximal) position
in the asynchronous game associated to the formula.
This game implements the interaction between a boolean function
(Proponent) of type (\ref{equation/b-tensor-b-pop-b})
and its two arguments (Opponent).
In a typical play, Opponent starts by playing the move~$\questionmove$
asking the value of the boolean output; Proponent reacts
by asking with $\questionmove_L$ the value of the left input,
and Opponent answers~$\truemove_L$;
then, Proponent asks with $\questionmove_R$ the value of the right input,
and Opponent answers~$\falsemove_R$; at this point only,
using the knowledge of its two arguments,
Proponent answers~$\falsemove$ to the initial question:
\begin{equation}
\label{equation/LR}
\questionmove \cdot \questionmove_L \cdot \truemove_L \cdot 
\questionmove_R \cdot \falsemove_R \cdot \falsemove
\end{equation}
Of course, Proponent could have explored its two arguments
in the other order, from right to left, this inducing the play
\begin{equation}
\label{equation/RL}
\questionmove \cdot \questionmove_R \cdot \falsemove_R \cdot 
\questionmove_L \cdot \truemove_L \cdot \falsemove
\end{equation}
The two plays start from the empty position~$\ast$ and reach the same position
of the asynchronous game.
They may be seen as different linearizations (in the sense of order theory)
of the partial order~(\ref{equation/slice-of-the-game})
provided by the game.
Each of these linearizations may be represented by adding causality (dotted) edges
between moves to the original partial order~(\ref{equation/slice-of-the-game}),
in the following way:
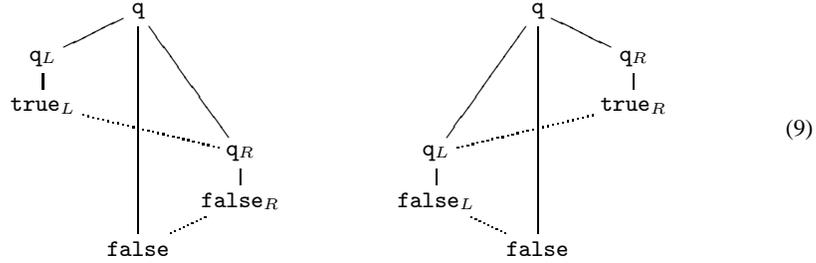
\begin{small}
\begin{equation}\label{equation/two-implementations}
\begin{array}{ccc}
\vcenter{\xymatrix @-1.5pc{
&& \questionmove\ar@{-}[dl]\ar@{-}[ddddd]\ar@{-}[dddr]&&
\\
& \questionmove_L \ar@{-}[d] && 
\\
& \truemove_L \ar@{.}[drr] && 
\\
&&& \questionmove_R \ar@{-}[d]
\\
&&& \falsemove_R \ar@{.}[dl]
\\
&& \falsemove &&
}}
&
\hspace{1em}
&
\vcenter{\xymatrix @-1.5pc{
&& \questionmove\ar@{-}[dr]\ar@{-}[ddddd]\ar@{-}[dddl]&&
\\
&&& \questionmove_R \ar@{-}[d] && 
\\
&&& \truemove_R \ar@{.}[dll] && 
\\
& \questionmove_L \ar@{-}[d]
\\
& \falsemove_L \ar@{.}[dr]
\\
&& \falsemove &&
}}
\end{array}
\end{equation}
\end{small}

\noindent
The play~(\ref{equation/LR}) is an element of the strategy
representing the \emph{left} implementation of the \emph{strict conjunction},
whereas the play~(\ref{equation/RL}) is an element of the strategy
representing its \emph{right} implementation.
Both of these strategies are alternating.
Now, there is also a \emph{parallel} implementation, where the conjunction
asks the value of its two arguments at the same time.
The associated strategy is not alternating anymore: it contains
the play~(\ref{equation/LR}) and the play~(\ref{equation/RL}),
and moreover, all the (possibly non-alternating) linearizations
of the following partial order:
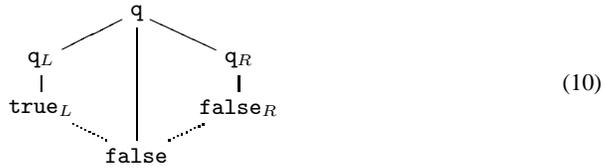
\begin{small}
\begin{equation}\label{equation/parallel-implementation}
\vcenter{\xymatrix @-1.5pc{
&& \questionmove\ar@{-}[dl]\ar@{-}[dr]\ar@{-}[ddd]&&
\\
& \questionmove_L \ar@{-}[d] && \questionmove_R \ar@{-}[d]
\\
& \truemove_L \ar@{.}[dr] && \falsemove_R \ar@{.}[dl]
\\
&& \falsemove &&
}}
\end{equation}
\end{small}

\noindent
This illustrates an interesting phenomenon, of a purely concurrent nature:
every play~$s$ of a concurrent strategy~$\sigma$ coexists
with other plays $t$ in the strategy, 
having the same target position~$x$ --
and in fact, equivalent modulo homotopy.
It is possible to reconstruct from this set of plays a \emph{partial order}
on the events of~$x$, refining the partial order on events provided by the game.
This partial order describes entirely the strategy~$\sigma$ under the position~$x$:
more precisely, the set of plays in~$\sigma$ reaching the position~$x$ coincides
with the set of the linearizations of the partial order.
Our definition of innocent strategy will ensure the existence of such an underlying
``causality order'' for every position~$x$ reached by the strategy.
Every innocent strategy will then define an event structure, 
obtained by putting together all the induced partial orders.
The construction requires refined tools from the theory of asynchronous transition systems,
and in particular the fundamental, but perhaps too confidential, notion of \emph{cube property}.

\myparagraph{Ingenuous strategies.}
We introduce in Section~\ref{section/ingenuous} the notion of \emph{ingenuous}
strategy, defined as a strategy regulated by an underlying ``causality order'' on moves
for every reached position, and satisfying a series of suitable diagrammatic properties.
One difficulty is that ingenuous strategies do not compose properly. 
Consider for instance the ingenuous strategy~$\sigma$
of type~$\booleangame\lltens\booleangame$
generated by the partial order:
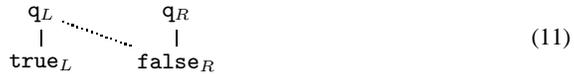
\begin{small}
\begin{equation}
\label{equation/counter-strategy}
\vcenter{\xymatrix @-1.5pc{
\questionmove_L \ar@{-}[d]\ar@{.}[drr] && \questionmove_R \ar@{-}[d]
\\
\truemove_L && \falsemove_R
}}
\end{equation}
\end{small}

\noindent
The strategy answers~$\truemove_L$ to the question~$\questionmove_L$,
but answers $\falsemove_R$ to the question~$\questionmove_R$
only if the question~$\questionmove_L$ has been already asked.
Composing the strategy~$\sigma$ with the \emph{right} implementation
of the strict conjunction pictured on the right handside of~(\ref{equation/two-implementations})
induces a play $\questionmove\cdot\questionmove_R$ stopped by
a \emph{deadlock} at the position~$\{\questionmove,\questionmove_R\}$.
On the other hand, composing the strategy with the \emph{left} or the
\emph{parallel} implementation is fine, and leads to a complete interaction.
This dynamic phenomenon is better understood by introducing
two new binary connectives~$\before$ and~$\after$ called ``before'' and ``after'',
describing sequential composition in asynchronous games.
The game~$A\before B$ is defined as the 2-dimensional restriction
of the game~$A\lltens B$ to the plays~$s$ such that every move played
before a move in~$A$ is also in~$A$; or equivalently, every move played
after a move~$B$ is also in~$B$.
The game~$A\after B$ is simply defined as the game~$B\before A$,
where the component~$B$ thus starts.
Now, the ingenuous strategy~$\sigma$ in~$\booleangame\lltens\booleangame$
specializes to a strategy in the subgame~$\booleangame\before\booleangame$,
which reflects it, in the sense that every play~$s\in\sigma$ is equivalent modulo homotopy
to a play~$t\in\sigma$ in the subgame~$\booleangame\before\booleangame$.
This is not true anymore when one specializes the strategy~$\sigma$
to the subgame~$\booleangame\after\booleangame$, because
the play~$\questionmove_L\cdot\truemove_L\cdot\questionmove_R\cdot\falsemove_R$
is an element of~$\sigma$ which is not equivalent modulo homotopy to any play~$t\in\sigma$
in the subgame~$\booleangame\after\booleangame$.
For that reason, we declare that the strategy~$\sigma$ is innocent 
in the game~$\booleangame\before\booleangame$
but \emph{not} in the game~$\booleangame\lltens\booleangame$.

\myparagraph{Innocent strategies.}
This leads to an interactive criterion which tests dynamically whether
an ingenuous strategy~$\sigma$ is \emph{innocent} for a given formula of linear logic.
The criterion is based on \emph{scheduling conditions} 
which recast, in the framework of non-alternating games,
the \emph{switching conditions} formulated by Abramsky and Jagadeesan
for alternating games~\cite{abramsky-jagadeesan:games-full-completeness}.
The idea is to \emph{switch} every tensor product~$\lltens$ of the formula
as $\before$ or $\after$ and to test whether every play~$s$ in the strategy~$\sigma$
is equivalent modulo homotopy to a play $t\in\sigma$ in the induced subgame.
Every such switching reflects a choice of scheduling by the counter-strategy:
an innocent strategy is thus a strategy flexible enough to adapt
to \emph{every} scheduling of the tensor products by Opponent.
An ingenuous strategy satisfies the scheduling criterion if and only if the underlying proof-structure
satisfies a directed (and more liberal) variant of the acyclicity criterion
introduced by Girard~\cite{girard:linear-logic}
and reformulated by Danos and Regnier~\cite{danos-regnier:multiplicatives}.
A refinement based on the notion of synchronized clusters of moves enables then to strengthen
the scheduling criterion, and to make it coincide with the usual non-directed acyclicity criterion.

We will establish in Section~\ref{section/ingenuous} that every ingenuous strategy
may be seen alternatively as a closure operator, whose fixpoints
are precisely the \emph{halting positions} of the strategy.
This connects (non alternating) asynchronous games to concurrent games.
However, there is a subtle mismatch between the interaction
of two ingenuous strategies seen as sets of plays, and seen
as sets of positions.
Typically, the right implementation of the strict conjunction
in~(\ref{equation/two-implementations}) composed
to the strategy~$\sigma$ in~(\ref{equation/counter-strategy})
induces two different fixpoints in the concurrent game model:
the deadlock position~$\{\questionmove,\questionmove_R\}$ reached
during the asynchronous interaction, and the complete position~$\{\questionmove,\questionmove_L,\truemove_L,\questionmove_R,\truemove_R, \falsemove\}$
which is never reached interactively.
The innocence assumption is precisely what ensures that this
will never occur: the fixpoint computed in the concurrent game model is unique,
and coincides with the position eventually reached in the asynchronous game model.
In particular, innocent strategies compose properly.

\myparagraph{Plan of the paper.}
We did our best to give in this introduction an informal but detailed overview
of this demanding work, which combines together ideas
from several fields: game semantics, concurrency theory, linear logic, etc.
We focus now on the conceptual properties of innocent strategies,
expressed in the diagrammatic language of asynchronous transition
systems.
The cube property is recalled in Section~\ref{section/cube}.
The diagrammatic properties of positionality are studied 
in Section~\ref{section/positional}.
The notion of ingenuous strategy is introduced in Section~\ref{section/ingenuous},
and reformulated as a class of well-behaved closure operators
in Section~\ref{section/concurrent-games}.
Finally, the notion of innocent strategy is defined in Section~\ref{section/innocence},
by strengthening the notion of ingenuous strategy with a scheduling criterion
capturing the essence of the acyclicity criterion of linear logic.

%


\section{The cube property}
\label{section/cube}
The \emph{cube property} expresses a fundamental causality principle
in the diagrammatic language of asynchronous transition
systems~\cite{bednarczyk:cas,shields:cm,winskel-nielsen:models-concur}.
The property is related to stability in the sense of Berry~\cite{berry:modeles-stables}.
It was first noticed by Nielsen, Plotkin and Winskel in~\cite{nielsen-plotkin-winskel:petri-event-domains},
then reappeared in~\cite{stabseq} and~\cite{mellies:ast,mellies:art1} and was studied thoroughly
by Kuske in his PhD thesis; see~\cite{droste-kuske:automata-concur} for a survey.
The most natural way to express the property is to start from
what we call an \emph{asynchronous graph}.
Recall that a \emph{graph} $G=(V,E,\partial_0,\partial_1)$ consists of a set $V$ of vertices
(or \emph{positions}), a set $E$ of edges (or \emph{transitions}),
and two functions $\partial_0,\partial_1:E\to V$ called respectively source and target functions.
An \emph{asynchronous graph} $\G=(G,\tile{})$ is a graph $G$ together
with a relation $\tile{}$ on coinitial and cofinal transition paths of length~2.
Every relation $s \tile{} t$ is represented diagrammatically as a 2-dimensional tile
\begin{equation}\label{equation/tuile2}
\vxym{
  &\ar[dl]_{m}x\ar[dr]^{n}&\\
  y_1\ar[dr]_{p}&\sim&\ar[dl]^{q}y_2\\
  &z&\\
}
\end{equation}
where $s=m\cdot p$ and $t=n\cdot q$.
In this diagram, the transition~$q$ is intuitively the \emph{residual}
of the transition~$m$ after the transition~$n$.
One requires the two following properties for every asynchronous tile:
\begin{enumerate}
\item $m\neq n$ and $p\neq q$,
\item the pair of transitions~$(n,q)$ is uniquely determined by the pair of transitions~$(m,p)$,
and conversely.
\end{enumerate}
%
%
The main difference with the asynchronous tile~(\ref{equation/tuile})
occurring in the asynchronous transition systems
defined in~\cite{winskel-nielsen:models-concur,sassone-nielsen-winskel:concurrency-classification}
is that the transitions are not labelled by events:
so, the 2-dimensional structure is purely ``geometric'' and
not deduced from an independence relation on events.
What matters is that the 2-dimensional structure enables one to define
a homotopy relation~$\sim$ on paths in exactly the same way.
%
%
Moreover, every homotopy class of a path~$s=m_1\cdots m_k$
coincides with the set of linearizations of a partial order on its transitions
if, and only if, the asynchronous graph satisfies the following \emph{cube property}:
\begin{quote}
\emph{Cube property:} a hexagonal diagram induced by two coinitial and cofinal
paths $m\cdot n\cdot o:x\transitionpath{} y$ and $p\cdot q\cdot r:x\transitionpath{} y$
is filled by 2-dimensional tiles as pictured in the lefthand side of the diagram below,
if and only it is filled by 2-dimensional tiles as pictured in the righthand side of the diagram:
\end{quote}
  \begin{center}
    $\vxym{
      &
      x\ar[dl]_m\ar[dd]\ar[rr]^p\hlbl{ddrr}
      &&
      x_2\ar[dd]^q
      \\
      x_1\ar[dd]_n\hlbl{dr}
      &&&
      \\
      &
      x_3 \ar[dl]\ar[rr]\hlbl{dr}
      &&
      y_1\ar[dl]^r
      \\
      y_2 \ar[rr]_o
      &&
      y
      &\\
    }\qquad\Longleftrightarrow\qquad
    \vxym{
      &
      x \ar[dl]_m\hlbl{dr}\ar[rr]^p
      &&
      x_2 \ar[dl]\ar[dd]^q
      \\
      x_1 \ar[dd]_n\hlbl{ddrr}\ar[rr]
      &&
      y_3 \hlbl{dr}\ar[dd]
      \\
      &&&
      y_1 \ar[dl]^r
      \\
      y_2 \ar[rr]_o
      &&
      y
      &
      \\
    }
    $
  \end{center}
The cube property is for instance satisfied by every asynchronous transition system
and every transition system with independence in the sense
of~\cite{winskel-nielsen:models-concur,sassone-nielsen-winskel:concurrency-classification}.
The correspondence between homotopy classes and sets of linearizations
of a partial order adapts, in our setting, a standard result on pomsets and
asynchronous transition systems with dynamic independence due to
Bracho, Droste and Kuske~\cite{bracho-droste-kuske:computations-concurrent-automata}.
Every asynchronous graph~$\G$ equipped with a distinguished initial position
(noted~$\ast$) induces an asynchronous graph~$[\G]$ whose positions
are the homotopy classes of paths starting from the position~$\ast$,
and whose edges $m:[s]\transition{}[t]$ between the homotopy classes
of the paths~$s:\ast\transitionpath{}x$ and~$t:\ast\transitionpath{}y$
are the edges~$m:x\transition{}y$ such that~$s\cdot m\sim t$.
When the original asynchronous graph~$\G$ satisfies the cube property,
the resulting asynchronous graph~$[\G]$ is ``contractible'' in the sense
that every two coinitial and cofinal paths are equivalent modulo homotopy.
So, we will suppose from now on that all our asynchronous graphs
satisfy the cube property and are therefore contractible.
The resulting framework is very similar to the domain of configurations
of an event structure.
Indeed, every contractible asynchronous graph defines a partial order
on its set of positions, defined by reachability: $x\leq y$ when $x\transitionpath{} y$.
Moreover, this order specializes to a finite distributive lattice under every position~$x$,
rephrasing -- by Birkhoff representation theorem -- the fact already mentioned
that the homotopy class of a path~$\ast\transitionpath{}x$ coincides
with the linearizations of a partial order on its transitions.
Finally, every transition may be labelled by an ``event'' representing the transition
modulo a ``zig-zag'' relation, identifying the moves~$m$ and~$q$ in every
asynchronous tile~(\ref{equation/tuile2}).
The idea of ``zig-zag'' is folklore: it appears for instance
in~\cite{sassone-nielsen-winskel:concurrency-classification}
in order to translate a transition system with independence
into a labelled event structure.
%

%


%

\section{Positionality in asynchronous games}
\label{section/positional}
Before considering 2-Player games, we express the notion
of a positional strategy in 1-Player games.
A 1-Player game~$(\G,\ast)$ is simply defined as an asynchronous graph~$\G$
together with a distinguished initial position~$\ast$.
A play is defined as a path starting from~$\ast$, and a strategy is defined
as a set of plays of the 1-Player game, closed under prefix.
A strategy~$\sigma$ is called \emph{positional} when for every three paths
$s,t:\initpos{}\transitionpath{}x$ and $u:x\transitionpath{}y$,
we have
\begin{equation}\label{equation/positionality-non-alternating}
s\cdot u \in \sigma
\hspace{1em}
\mbox{and}
\hspace{1em}
s\sim t
\hspace{1em}
\mbox{and}
\hspace{1em}
t \in \sigma
\hspace{1em}
\text{implies}
\hspace{1em}
t\cdot u \in \sigma.
\end{equation}
This adapts the definition of~(\ref{equation/positionality-alternating})
to a non-polarized setting and extends to the non-alternating setting.
Note that a positional strategy is the same thing as a subgraph of the 1-Player game,
where every position is reachable from~$\ast$ inside the subgraph.
This subgraph inherits a 2-dimensional structure from the underlying 1-Player game;
it thus defines an asynchronous graph, denoted~$\G_\sigma$.
The advantage of considering asynchronous graphs instead of event structures
appears at this point: the ``event'' associated to a transition is deduced
from the 2-dimensional geometry of the graph.
%
%
So, the ``event'' associated to a transition~$m$ in the graph~$\G_\sigma$
describes the ``causality cascade'' leading the strategy to play the transition~$m$ ;
whereas the ``event'' associated to the same transition~$m$
in the 1-Player game~$\G$ is simply the move of the game.
This subtle difference is precisely what underlies the distinction between
the formula~(\ref{equation/b-tensor-b-pop-b}) and the various
strategies~(\ref{equation/two-implementations})
and~(\ref{equation/parallel-implementation}).
For instance, there are three ``events'' associated to the output move~$\falsemove$
in the parallel implementation of the strict conjunction, each one corresponding
to a particular pair of inputs $(\truemove,\falsemove)$, $(\falsemove,\falsemove)$,
and $(\falsemove,\truemove)$.
This phenomenon is an avatar of Berry stability, already noticed in~\cite{mellies:art4}.
From now on, we only consider positional strategies
satisfying two additional properties:
\begin{enumerate}
\item forward compatibility preservation: every asynchronous tile
of the shape~(\ref{equation/tuile2}) in the 1-Player game~$\G$
belongs to the subgraph~$\G_\sigma$ of the strategy~$\sigma$
when its two coinitial transitions $m:x\transition{}y_1$ and $n:x\transition{}y_2$
are transitions in the subgraph~$\G_\sigma$. Diagrammatically,
$$
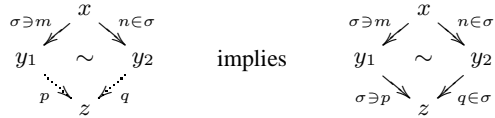
\vxym{
  &\ar[dl]_{\sigma\owns m}x\ar[dr]^{n\in\sigma}&\\
  y_1\ar@{.>}[dr]_{p}&\sim&\ar@{.>}[dl]^{q}y_2\\
  &z&\\
}
\qqtimpl
\vxym{
  &\ar[dl]_{\sigma\owns m}x\ar[dr]^{n\in\sigma}&\\
  y_1\ar[dr]_{\sigma\owns p}&\sim&\ar[dl]^{q\in\sigma}y_2\\
  &z&\\
}
$$
where the dotted edges indicate edges in $\G$.
\item backward compatibility preservation:
dually, every asynchronous tile of the shape~(\ref{equation/tuile2}) in the 1-Player game~$\G$
belongs to the subgraph~$\G_\sigma$ of the strategy~$\sigma$
when its two cofinal transitions $p:y_1\transition{}z$ and $q:y_2\transition{}z$
are transitions in the subgraph~$\G_\sigma$.
\end{enumerate}
These two properties ensure that the asynchronous graph~$\G_\sigma$ 
is contractible and satisfies the cube property.
Contractibility means that every two cofinal plays $s,t:\ast\transitionpath{}x$ of
the strategy~$\sigma$ are equivalent modulo homotopy \emph{inside}
the asynchronous graph~$\G_\sigma$ -- that is, every intermediate play
in the homotopy relation is an element of~$\sigma$.
Moreover, there is a simple reformulation as a set of plays
of a positional strategy satisfying the two preservation properties: it is (essentially) a set of plays
satisfying (1) a suitable cube property and (2)
that $s\cdot m\in\sigma$ and $s\cdot n\in\sigma$ implies 
$s\cdot m\cdot p\in\sigma$ and $s\cdot n\cdot q\in\sigma$
when~$m$ and~$n$ are the coinitial moves of a tile~(\ref{equation/tuile2}).
This characterization enables us to regard a positional strategy
either as a set of plays, or as an asynchronous subgraph of the game.

\section{Ingenuous strategies in asynchronous games}
\label{section/ingenuous}
A 2-Player game $(\G,\ast,\lambda)$ is defined as an asynchronous graph $\G=(V,E,\tile{})$
together with a distinguished initial position~$\ast$, and a function $\lambda:E\to\{-1,+1\}$
which associates a \emph{polarity} to every transition (or move) of the graph.
Moreover, the equalities~$\lambda(m)=\lambda(q)$ and~$\lambda(n)=\lambda(p)$
are required to hold in every asynchronous tile~(\ref{equation/tuile2})
of the asynchronous graph~$\G$.
The convention is that a move~$m$ is played by \emph{Proponent}
when $\lambda(m)=+1$ and by \emph{Opponent} when $\lambda(m)=-1$.
A strategy~$\sigma$ is called \emph{ingenuous} when it satisfies the following properties:
\begin{enumerate}
  \item it is \emph{positional}, and satisfies the backward and forward compatibility
  preservation properties of Section~\ref{section/positional},
  \item it is \emph{deterministic}, in the following concurrent sense:
  every pair of coinitial moves~$m:x\transition{}y_1$ and~$n:x\transition{}y_2$
  in the strategy~$\sigma$ where the move~$m$ is played by Proponent,
  induces an asynchronous tile~(\ref{equation/tuile2}) in the strategy~$\sigma$.
  Diagrammatically,
  $$
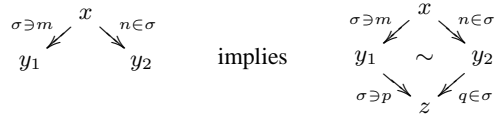
\vxym{
  &\ar[dl]_{\sigma\owns m}x\ar[dr]^{n\in\sigma}&\\
  y_1\ar@{}[dr]&&\ar@{}[dl]y_2\\
  &&\\
}
\qqtimpl
\vxym{
  &\ar[dl]_{\sigma\owns m}x\ar[dr]^{n\in\sigma}&\\
  y_1\ar[dr]_{\sigma\owns p}&\sim&\ar[dl]^{q\in\sigma}y_2\\
  &z&\\
}
$$
  \item it is \emph{courteous}, in the following sense: 
  every asynchronous tile~(\ref{equation/tuile2}) where the two moves
 $m:x\transition{}y_1$ and $p:y_1\transition{}z$ are in the strategy~$\sigma$,
  and where $m$ is a Proponent move,
  is an asynchronous tile in the strategy~$\sigma$.
  Diagrammatically,
$$
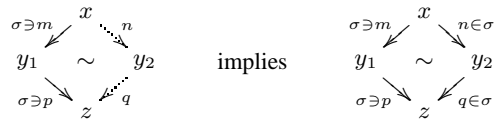
\vxym{
  &\ar[dl]_{\sigma\owns m}x\ar@{.>}[dr]^{n}&\\
  y_1\ar[dr]_{\sigma\owns p}&\sim&\ar@{.>}[dl]^{q}y_2\\
  &z&\\
}
\qqtimpl
\vxym{
  &\ar[dl]_{\sigma\owns m}x\ar[dr]^{n\in\sigma}&\\
  y_1\ar[dr]_{\sigma\owns p}&\sim&\ar[dl]^{q\in\sigma}y_2\\
  &z&\\
}
$$
when $\lambda(m)=+1$.
\end{enumerate}
Note that, for simplicity, we express this series of conditions on strategies
seen as asynchronous subgraphs. However, the conditions may be reformulated
in a straightforward fashion on strategies defined as sets of plays.
The forward and backward compatibility preservation properties
of Section~\ref{section/positional} ensure that the set of plays
of the strategy~$\sigma$ reaching the same position~$x$
is regulated by a ``causality order'' on the moves occurring
in these plays -- which refines the ``justification order'' on moves
(in the sense of pointer games) provided by the asynchronous game.

Our concurrent notion of determinism is not exactly the same
as the usual notion of determinism in sequential games: in particular,
a strategy may play several Proponent moves from a given position,
as long as it converges later.
Courtesy ensures that a strategy~$\sigma$ which accepts an Opponent move~$n$
\emph{after} playing an independent Proponent move~$m$, is ready to delay
its own action, and to accept the move~$n$ \emph{before} playing the move~$m$.
Together with the receptivity property introduced in Section~\ref{section/innocence},
this ensures that the ``causality order'' on moves induced
by such a strategy refines the underlying ``justification order''
of the game, by adding only order dependencies~$m\preceq n$
where $m$ is an Opponent move and $n$ is a Proponent move.
This adapts to the non-alternating setting the fact that, in alternating games,
the causality order~$p\preceq q$ provided by the view of an innocent strategy
coincides with the justification order when~$p$ is Proponent and~$q$ is Opponent.

The experienced reader will notice that an ingenuous (and not necessarily receptive)
strategy may add order dependencies~$m\preceq n$ between two Opponent moves~$m$ and~$n$.
In the next Section, we will see that this reflects an unexpected property
of concurrent strategies expressed as closure operators.

\section{Ingenuous strategies in concurrent games}
\label{section/concurrent-games}
In this section, we reformulate ingenuous strategies in asynchronous games
as strategies in concurrent games.
The assumption that our 2-Player games are played on contractible
asynchronous graphs induces a partial order on the set of positions,
defined by reachability: $x\leq y$ when $x\transitionpath{}y$.
The concurrent game associated to an asynchronous game~$\G$
is defined as the ideal completion of the partial order of positions reachable
(in a finite number of steps) from the root~$\ast$.
So, the positions in the concurrent game are either \emph{finite}
when they are reachable from the root, or \emph{infinite} when they
are defined as an infinite directed subset of finite positions.
The complete lattice~$D$ is then obtained by adding a top element~$\top$
to the ideal completion.
Considering infinite as well as finite positions introduces technicalities
that we must confront in order to cope with infinite interactions,
and to establish the functoriality property at the end of Section~\ref{section/innocence}.
Now, we will reformulate ingenuous strategies in the asynchronous game~$\G$
as \emph{continuous} closure operators on the complete lattice~$D$.
By continuous, we mean that the closure operator preserves joins of directed subsets.
Every closure operator~$\sigma$ on a complete lattice~$D$ induces a set of fixpoints:
\begin{equation}
\label{equation/fixpoints-from-closure}
\fixpoints{\sigma} \qeq \{ \hspace{.4em} x\in D \tq \sigma(x)=x  \hspace{.4em}\}
\end{equation}
closed under arbitrary meets.
Moreover, when the closure operator~$\sigma$ is continuous,
the set~$\fixpoints{\sigma}$ is closed under joins of directed subsets.
Conversely, every subset~$X$ of the complete lattice~$D$
closed under arbitrary meets defines a closure operator
\begin{equation}
\label{equation/closure-from-fixpoints}
\sigma \qcolon x 
\hspace{.4em}
\mapsto
\hspace{.4em}
\bigwedge
\hspace{.4em}
 \{
 \hspace{.4em}
 y\in X \tq x\leq y
 \hspace{.4em}
 \}
\end{equation}
which is continuous when the subset~$X$ is closed under joins of directed subsets.
Moreover, the two translations~(\ref{equation/fixpoints-from-closure})
and~(\ref{equation/closure-from-fixpoints}) are inverse operations.

Now, every ingenuous strategy~$\sigma$ defines a set~$\halting{\sigma}$
of \emph{halting positions}.
We say that a \emph{finite} position~$x$ is \emph{halting}
when (1) the position is reached by the strategy~$\sigma$ and
(2) there is no Proponent move~$m:x\transition{}y$ in the strategy~$\sigma$.
The definition of a halting position can be extended to infinite positions
by ideal completion: infinite positions are thus defined as downward-closed
directed subsets~$\hat{x}$ of finite positions.
We do not provide the details here for lack of space.
%
%
It can be shown that the set of halting positions of an ingenuous
strategy~$\sigma$ is closed under arbitrary meets, and under joins of directed subsets.
It thus defines a continuous closure operator, noted~$\sigma^\circ$,
defined by~(\ref{equation/closure-from-fixpoints}).
The closure operator~$\sigma^{\circ}$ satisfies a series of additional properties:
\begin{enumerate}
  \item The domain~$\domain{\sigma^\circ}$ is closed under (arbitrary) compatible joins,
  \item For every pair of positions $x,y\in\domain{\sigma^\circ}$
  such that~$x\leq y$, either~$\sigma^\circ(x)=\sigma^\circ(y)$
  or there exists an Opponent move~$m:\sigma^\circ(x)\transition{}z$
  such that~$z\leq_P \sigma^\circ(z)$ and $\sigma^\circ(z) \leq \sigma^{\circ}(y)$.
\end{enumerate}
Here, the \emph{domain}~$\domain{\sigma^{\circ}}$ of the closure operator~$\sigma^\circ$
is defined as the set of positions~$x\in D$ such that~$\sigma^\circ(x)\neq\top$;
and the Proponent reachability order~$\leq_P$ refines the reachability order~$\leq$
by declaring that $x\leq_P y$ means, for two finite positions~$x$ and~$y$,
that there exists a path $x\transitionpath{}y$ containing only Proponent moves;
and then, by extending the definition of~$\leq_P$ to all positions (either finite or infinite) in~$D$
by ideal completion.
Conversely, every continuous closure operator~$\tau$ which satisfies the two
additional properties mentioned above induces an ingenuous strategy~$\sigma$
in the following way.
The \emph{dynamic domain}
of the closure operator~$\tau$ is defined
as the set of positions~$x\in\domain{\tau}$ such that~$x\leq_P\tau(x)$.
So, a position~$x$ is in the dynamic domain of~$\tau$ when
the closure operator increases it in the proper way, that is,
without using any Opponent move.
The ingenuous strategy~$\sigma$ induced by the closure operator~$\tau$
is then defined as the set of plays
whose intermediate positions
are all in the dynamic domain of~$\tau$.
This defines an inverse to the operation $\sigma\mapsto \sigma^\circ$
from ingenuous strategies to continuous closure operators satisfying
the additional properties 1. and 2.
This pair of constructions thus provides a one-to-one correspondence between
the ingenuous strategies and continuous closure operators
satisfying the additional properties 1. and 2.

\section{Innocent strategies}
\label{section/innocence}
Despite the one-to-one correspondence between ingenuous strategies
and concurrent strategies described in Section~\ref{section/concurrent-games},
there is a subtle mismatch between the two notions -- which fortunately disappears
when ingenuity is refined into innocence.
On the one hand, it is possible to construct a category~$\mathcal{G}$ of asynchronous games
and ingenuous strategies, where composition is defined
by ``parallel composition+hiding'' on strategies seen as sets of plays.
%
%
On the other hand, it is possible to construct a category~$\mathcal{C}$ of concurrent games
and concurrent strategies, defined in~\cite{abramsky-mellies:concurrent-games},
where composition coincides with relational composition on the sets of fixpoints
of closure operators.
Unfortunately -- and here comes the mismatch -- the translation~$\mathcal{G}\rightarrow\mathcal{C}$
described in Section~\ref{section/concurrent-games}
is not functorial, in the sense that it does not preserve composition of strategies.
This is nicely illustrated by the example of the ingenuous strategy~(\ref{equation/counter-strategy}) 
whose composition with the right implementation of
the strict conjunction~(\ref{equation/two-implementations}) induces a deadlock.
More conceptually, this phenomenon comes from the fact that the category~$\mathcal{G}$
is compact closed: the tensor product~$\lltens$ is identified with its dual~$\llpar$.
%


%
This motivates a strengthening ingenuous strategies by a \emph{scheduling criterion}
which distinguishes the tensor product from its dual, and plays the role, in the non-alternating setting,
of the \emph{switching conditions} introduced by Abramsky and Jagadeesan
for alternating games~\cite{abramsky-jagadeesan:games-full-completeness}.
The criterion is sufficient to ensure that strategies do not deadlock during composition.
%
In order to explain it here, we limit ourselves, for simplicity, to formulas of multiplicative linear logic
(thus constructed using $\lltens$ and $\llpar$ and their units~$1$ and~$\bot$)
extended with the two lifting modalities~$\uparrow$ and $\downarrow$.
The tensor product, as well as its dual, are interpreted by the expected
``asynchronous product'' of asynchronous graphs:
in particular, every play~$s$ of~$A\lltens B$ may be seen as a pair of plays~$(s_A,s_B)$
of~$A$ and~$B$ modulo homotopy.
The two connectives~$\lltens$ and~$\llpar$ are then distinguished by attaching
a label~$\lltens$ or~$\llpar$ to every asynchronous tile~(\ref{equation/tuile2})
appearing in the game.
Typically, an asynchronous tile~(\ref{equation/tuile2}) between a move~$m$
in~$A$ and a move~$n$ in~$B$ is labelled by~$\lltens$ in the asynchronous
game~$A\lltens B$ and labelled by~$\llpar$ in the asynchronous game $A\llpar B$.
The lifting modality~$\uparrow$ (resp.~$\downarrow$) is then interpreted
as the operation of ``lifting'' a game with an initial Opponent
(resp. Proponent) move.
Note that there is a one-to-one relationship between the lifting modalities~$\uparrow$
and~$\downarrow$ appearing in the formula, and the moves of the asynchronous
game~$G$ which denotes the formula.
A nice aspect of our asynchronous approach is that we are able to formulate
our scheduling criterion in two alternative but equivalent ways, each of them capturing
a particular point of view on the correctness of proofs and strategies:
\begin{enumerate}
\item \textbf{a scheduling criterion} based on a switching 
as ``before''~$\before$ or ``after''~$\after$ of every tensor product~$\lltens$
in the underlying formula of linear logic.
As explained in the introduction, the scheduling criterion requires
that every path~$s$ in the strategy~$\sigma$ is equivalent modulo homotopy
to a path~$t$ in the strategy~$\sigma$ which respects the scheduling
indicated by the switching.
This is captured diagrammatically by orienting every~$\lltens$-tile
as~$\before$ or~$\after$ according to the switching, and by requiring
that every play~$s\in\sigma$ \emph{normalizes} to a 2-dimensional
normal form $t\in\sigma$ w.r.t. these semi-commutations~\cite{clerbout-latteux-roos}
or standardization tiles~\cite{mellies:art1}.
\item \textbf{a directed acyclicity criterion} which reformulates
the previous scheduling criterion along the lines of Girard's long trip criterion~\cite{girard:linear-logic}
and Danos-Regnier's acyclicity criterion~\cite{danos-regnier:multiplicatives}.
Every position~$x$ reached by an ingenuous strategy~$\sigma$
induces a partial order~$\preceq$ on the moves appearing in the position~$x$.
Every relation~$m\preceq n$ of the partial order induces a \emph{jump}
between the lifting modalities associated to the moves~$m$ and~$n$
in the formula.
The acyclicity criterion then requires that every switching of the $\llpar$ connectives
as ``Left'' or ``Right'' (that is, in the sense of Girard) induces a graph with no \emph{directed} cycles.
\end{enumerate}
The scheduling criterion ensures that the operation~$\sigma\mapsto\sigma^{\circ}$
defines a \emph{lax} functor, in the sense that every fixpoint of~$\sigma^\circ;\tau^\circ$
is also a fixpoint of $(\sigma;\tau)^\circ.$
Now, an ingenuous strategy~$\sigma$ is called \emph{asynchronous} when
it additionally satisfies the following \emph{receptivity} property:
for every play~$s:\ast\transitionpath{}x$ and for every move~$m:x\transition{}y$,
$$
s\in\sigma \hspace{1em} \mbox{and} \hspace{1em}  \mbox{$\lambda(m)=-1$}
\hspace{1em} \timpl \hspace{1em}  s\cdot m\in\sigma.
$$
The category~$\mathcal{A}$ is then defined as follows:
its objects are the asynchronous games equipped with $\lltens$-tiles and $\llpar$-tiles,
and its morphisms $A\rightarrow B$ are the asynchronous strategies of $A\llimp B$, defined as $A^\ast\llpar B$,
satisfying the scheduling criterion -- where~$A^\ast$ is the asynchronous game~$A$ with Opponent
and Proponent interchanged.
The scheduling criterion ensures that the operation~$\sigma\mapsto\sigma^{\circ}$ defines
a strong monoidal functor~$\mathcal{A}\rightarrow\mathcal{C}$ from the category~$\mathcal{A}$
to the category~$\mathcal{C}$ of concurrent games and concurrent strategies -- thus extending
the programme of~\cite{baillot:timeless,mellies:ag2} in the non-alternating setting.
The notion of asynchronous strategy is too liberal to capture the notion of innocent strategy,
at least because there exist asynchronous strategies which are not definable
as the interpretation of a proof of MLL extended with lifting modalities~$\uparrow$ and $\downarrow$.
The reason is that the scheduling criterion tests only for \emph{directed} cycles,
instead of the usual non-directed cycles.
On the other hand, it should be noted that the directed acyclicity criterion coincides with
the usual non-directed acyclicity criterion in the situation
treated in~\cite{abramsky-jagadeesan:games-full-completeness}
-- that is, when the formula is purely multiplicative (i.e. contains no lifting modality),
every variable~$X$ and~$X^\bot$ is interpreted as a game with a Proponent and an Opponent move,
and every axiom link is interpreted as a ``bidirectional'' copycat strategy.
The full completeness result in~\cite{abramsky-mellies:concurrent-games}
uses a similar directed acyclicity criterion for MALL.
Hence, directed acyclicity is a fundamental, but somewhat hidden, concept of game semantics.

On the other hand, we would like to mirror the usual non-directed acyclicity criterion
in our asynchronous and interactive framework.
This leads us to another stronger scheduling criterion based on the idea that
in every play~$s$ played by an asynchronous strategy~$\sigma$,
an Opponent move~$m$ and a Proponent move~$n$ appearing in the play,
and directly related by the causality order~$m\preceq n$ induced by~$\sigma$
can be played synchronously.
%
We write~$m\leftrightharpoons n$ in that case, and say that two moves
are synchronized in~$s$ when they lie in the same equivalence class generated
by $\leftrightharpoons$.
A cluster of moves $t$ in the play~$s=m_1\cdots m_k$ is then defined as a path
$t=m_i\cdots m_{i+j}$ such that all the moves appearing in~$t$ are synchronized.
Every play~$s$ in the strategy~$\sigma$ can be reorganized as a sequence of maximal
clusters, using a standardisation mechanism~\cite{mellies:art1,mellies:art4}.
The resulting clustered play is unique, modulo permutation of clusters, noted~$\sim_{OP}$.
This relation generalizes to the non-alternating case the relation~$\sim_{OP}$
introduced in~\cite{mellies:ag2}.
This leads to
\begin{itemize}
\item[3.] \textbf{a clustered scheduling criterion} based, just as previously,
on a switching as ``before''~$\before$ or ``after''~$\after$ of every tensor product~$\lltens$
in the underlying formula of linear logic.
The difference is that we ask that every clustered play~$s$ in the strategy~$\sigma$
may be reorganized modulo~$\sim_{OP}$ as a clustered play which respects the scheduling
indicated by the switching.
\end{itemize}
An asynchronous strategy is called innocent when it satisfies this stricter scheduling criterion.
Although this tentative definition of innocence is fine conceptually, 
we believe that it has to be supported by further proof-theoretic investigations.
An interesting aspect of our scheduling criteria is that they may be formulated
in a purely diagrammatic and 2-dimensional way: in particular, the switching conditions
are expressed here using the underlying logic MLL with lifting modalities~$\uparrow$
and $\downarrow$ for clarity only, and may be easily reformulated diagrammatically.



\section{Conclusion}
Extending the framework of asynchronous games to non-alternating strategies
requires an exploration of the fine-grained structure of causality, using classical concepts
of concurrency theory like the cube property.
Interestingly, it appears that enforcing good causality properties on strategies
is not sufficient to combine game semantics and concurrency theory in a harmonious way.
Indeed, we uncover a subtle and unexpected mismatch between composition
performed in asynchronous games and composition performed in concurrent games.
The mismatch is resolved by strengthening the purely causal notion
of \emph{ingenuous strategy} into the more contextual notion of \emph{innocent strategy}
by imposing a \emph{scheduling criterion} which reformulates in a purely
interactive and diagrammatic fashion the usual acyclicity criterion of linear logic.
The criterion is sufficient to ensure the existence of a strong monoidal functor
from the category of asynchronous games to the category of concurrent games.
%

%
%
%
%

%
In the near future, we plan to investigate the relationship between asynchronous games
and L-nets. The scheduling criterion should help, since L-nets are themselves based
on an acyclicity criterion~\cite{curien-faggian:L-nets}.
We also plan to investigate the relationship between asynchronous games and 
the recent work by Varacca and Yoshida on confusion-free event structures
and the $\pi$-calculus~\cite{varacca-yoshida:typed-event-structures}.
Our point of view that every strategy~$\sigma$ is an asynchronous graph~$G_\sigma$
embedded in its asynchronous game~$G$ is certainly a propitious starting point,
since it enables a study of the formal properties of these embeddings,
in the spirit of Nielsen and Winskel~\cite{winskel-nielsen:models-concur},
which precisely underlies the construction in~\cite{varacca-yoshida:typed-event-structures}.
These connections would support our claim that asynchronous games provide indeed
a valuable foundation for concurrency in game semantics.


\paragraph{Acknowledgments.}
We would like to thank Martin Hyland together with Pierre-Louis Curien, Claudia Faggian,
Russ Harmer, Daniele Varacca, and Nobuko Yoshida for spontaneous and lively
blackboard discussions.

\bibliographystyle{plain}
\bibliography{these}

\end{document}